\documentclass[twocolumn]{el-author}

\newcommand{\ua}{\uparrow}
\newcommand{\nc}{\newcommand}
\nc{\da}{\downarrow} \nc{\hc}{\hat{c}} \nc{\hS}{\hat{S}}
\nc{\bra}{\langle} \nc{\ket}{\rangle} \nc{\eq}{equation (\ref}
\nc{\h}{\hat} \nc{\hT}{\h{T}}\nc{\be}{\begin{eqnarray}}
\nc{\ee}{\end{eqnarray}}\nc{\rd}{\textrm{d}}\nc{\e}{eqnarray}\nc{\hR}{\hat{R}}\nc{\Tr}{\mathrm{Tr}}
\nc{\tS}{\tilde{S}}\nc{\tr}{\mathrm{tr}}\nc{\8}{\infty}\nc{\lgs}{\bra\ua,\phi|}\nc{\rgs}{|\ua,\phi\ket}
\nc{\hU}{\hat{U}}\nc{\lfs}{\bra\phi|}\nc{\rfs}{|\phi\ket}\nc{\hZ}{\hat{Z}}\nc{\hd}{\hat{d}}\nc{\mD}{\mathcal{D}}
\nc{\bd}{\bar{d}}\nc{\bc}{\bar{c}}\nc{\mc}{\mathcal}\nc{\ea}{eqnarray}\nc{\mG}{\mathcal{G}}\nc{\bce}{\begin{center}}
\nc{\ece}{\end{center}}
\date{17th June 2014}

\usepackage{subfigure}

\begin{document}

\title{Quasi-optimal grouping for broadcast systems with hierarchical modulation}
\author{Hugo M{\'e}ric, Jos{\'e} Miguel Piquer and J{\'e}r{\^o}me Lacan}

\abstract{Recently, we proposed to combine time sharing with hierarchical modulation to increase the transmission rate of broadcast systems. Our proposal involves to group the receivers in pairs in order to transmit with hierarchical modulation. We introduced several grouping strategies but the optimal matching remained an open question. In this letter, we show that the optimal grouping is the solution of an assignment problem, for which efficient algorithms exist such as the Hungarian method. Based on this algorithm, we study the performance of the optimal grouping in terms of spectrum efficiency for a DVB-S2 system.}

\maketitle

\section{Introduction}
Modern broadcasting systems such as DVB-S2\footnote{Digital Video Broadcasting - Satellite - Second Generation} or DVB-SH\footnote{Digital Video Broadcasting - Satellite to Handheld} rely mainly on a time sharing strategy to optimise the transmission rate. Recently, we showed that combining time sharing with hierarchical modulation, a technique that merges several data streams in a same symbol, can provide significant gains (in terms of spectrum efficiency) compared to the best time sharing strategy \cite{1,2}.

In this letter, we consider one source communicating with $n$ receivers. The objective is to offer the \emph{same average spectrum efficiency to all the receivers}. We assume that the transmitter has knowledge of the signal-to-noise ratio (SNR) at the receivers. A concrete example is a DVB-S2 system that implements adaptive coding and modulation (ACM). Moreover, the system also implements \emph{hierarchical modulation with two layers}, i.e., two data streams are merged in a same symbol. 

Receiver $i$ ($1 \leqslant i \leqslant n$) has a spectrum efficiency $R_i$ which corresponds to the best spectrum efficiency it can manage. The value of $R_i$ depends on $SNR_i$, the SNR of receiver $i$, and the transmission parameters (code rate and modulation). For instance, if the source transmits a QPSK modulated signal and the code rate is 1/3, then $R_i = 2\times 1/3$ bit/symbol. However, if receiver $i$ is paired with receiver $j$ ($1 \leqslant j \leqslant n$, $j \neq i$) and hierarchical modulation is used, the spectrum efficiency of receivers $i$ and $j$ is $R_{ij}^{hm}$ which depends on $SNR_i$, $SNR_j$ and the transmission parameters. The computation of $R_{ij}^{hm}$ is detailed in \cite{1}.

\begin{figure}[h]
\centering
\includegraphics[width=0.7\columnwidth]{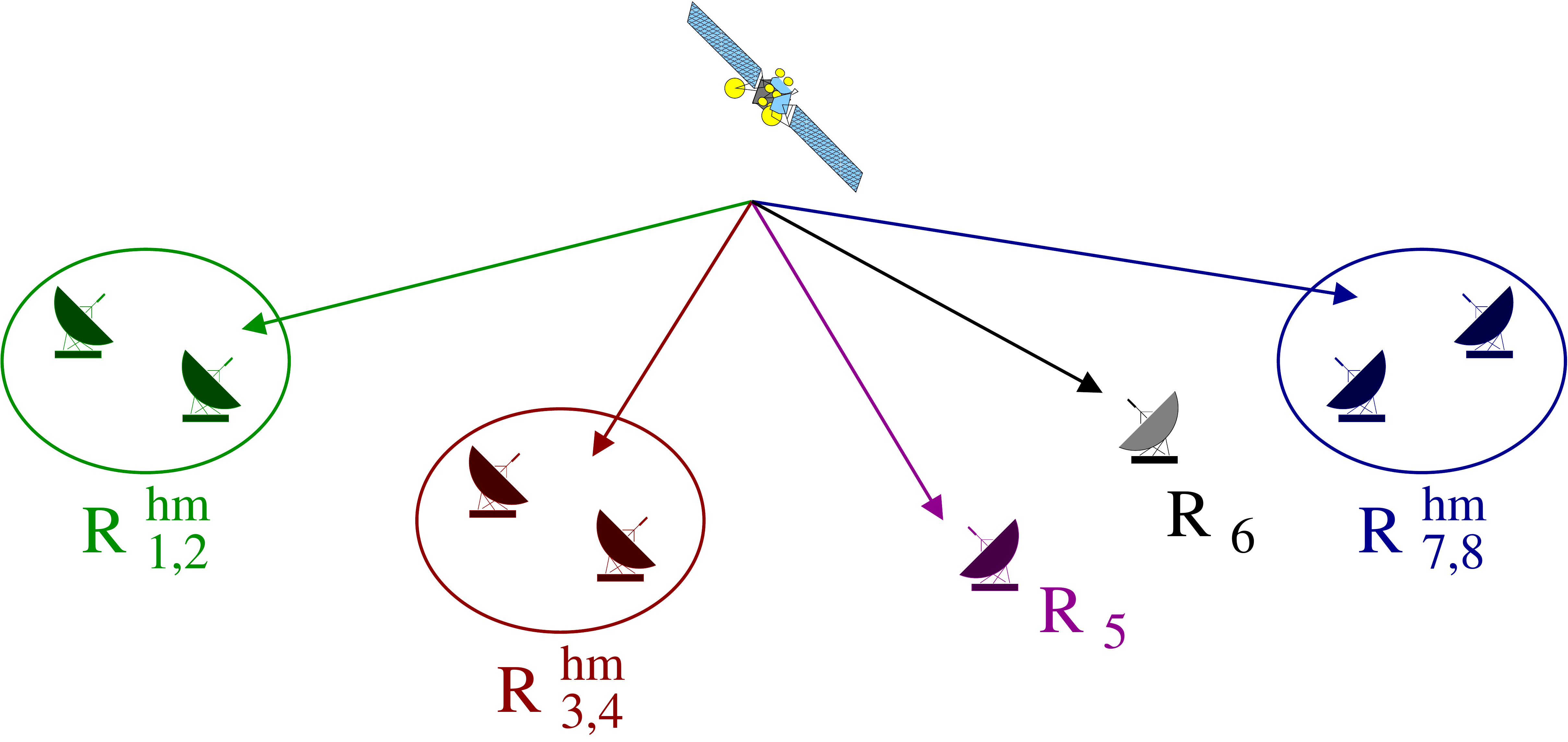}
\caption{Broadcast system with 8 receivers and a given grouping strategy}
\label{hm}
\end{figure}

During the transmission, the source can either communicate directly with a receiver (called a single receiver) or group it with another receiver and use hierarchical modulation (called paired receivers). This process is the grouping strategy or matching. Once the strategy is decided, the average spectrum efficiency offered to all the receivers is
\begin{equation}
R = \left( \sum_{k} \frac{1}{R_{k}} + \sum_{(i,j)} \frac{1}{R^{hm}_{ij}} \right)^{-1},
\label{rate}
\end{equation}
where the sum over $k$ takes into account the single receivers and the sum over $(i,j)$ the paired receivers. Equation (\ref{rate}) is a direct extension of (9) in \cite{2}. As an example, \figurename~\ref{hm} illustrates a system with 8 receivers and a given grouping strategy. In that case, the average spectrum efficiency is
\begin{equation}
R = \left( \frac{1}{R^{hm}_{1,2}} + \frac{1}{R^{hm}_{3,4}} + \frac{1}{R_5} + \frac{1}{R_6} + \frac{1}{R^{hm}_{7,8}} \right)^{-1}.
\end{equation}

Compared to our previous works \cite{1,2}, the matching considered in this letter allows to communicate directly with a receiver without pairing it with another receiver. Thus the framework is more general.

\section{Optimal matching}
In \cite{1}, we introduced several grouping strategies. Among the proposed strategies, the one that achieves the best performance consists in grouping the two receivers with the largest SNR difference and repeating this operation until each receiver is in a pair. However, the optimal matching, i.e., the one that maximises the average spectrum efficiency in (\ref{rate}), remained an open question.

For a system with $n$ receivers, we note $s_n$ the number of possible strategies consisting in single or paired receivers. The sequence $(s_n)_{n \geqslant 1}$ verifies
\begin{equation}
s_n = s_{n-1} + (n-1) s_{n-2},
\label{sn}
\end{equation}
for $n \geqslant 3$, with $s_1=1$ and $s_2=2$. To obtain (\ref{sn}), we consider a system with $n$ receivers. Then receiver $n$ can either be a single receiver and in that case we group the remaining $n-1$ receivers; or  receiver $n$ can be paired with another receiver (there are exactly $n-1$ possibilities) and we group the remaining $n-2$ receivers. From (\ref{sn}), we can show by recursion that $s_k \geqslant 2^k$ for $k \geqslant 5$. For large broadcast systems, it is thus impossible to test all the possible matchings to determine the optimal one. 

To obtain the optimal strategy, we write the problem in matrix form. A grouping strategy can be represented by a $n \times n$ \emph{assignment matrix} $X$ where
\begin{equation}
X_{i,j} =  \begin{cases} 1 , & \text{if receiver } i \text{ is paired with receiver } j \\ 0 , & \text{otherwise.}\end{cases}
\end{equation}
By definition, $X$ is symmetric, contains exactly $n$ non-zero entries and there is exactly one non-zero entry in each row and column. In other words, $X$ is a symmetric permutation matrix. Note that the ones in the diagonal correspond to the single receivers. Then, we define the $n \times n$ \emph{cost matrix} $C$ by
\begin{equation}
C_{i,j} =  \begin{cases} 1/R_i , & \text{if } i=j \\ 1/\left(2R_{ij}^{hm}\right) , & \text{if } i \neq j \end{cases}
\label{cost}
\end{equation}
where $1 \leqslant i,j \leqslant n$. The cost matrix is also symmetric. 

In combinatorial optimisation, the \emph{assignment problem} can be formulated as follow: given a cost matrix $C$, find an assignment (i.e., a set of $n$ entry positions so no two of which lie in the same row or column) such that the sum of the $n$ entries is the smallest possible. The sum of the $n$ entries is the \emph{assignment cost}. By noting that maximising the average spectrum efficiency in (\ref{rate}) amounts to minimising the term
\begin{equation}
\sum_{k} \frac{1}{R_{k}} + \sum_{(i,j)} \frac{1}{R^{hm}_{ij}},
\label{term_to_minimise}
\end{equation}
the optimal grouping strategy is (almost) equivalent to an assignment problem with the cost matrix $C$ defined in (\ref{cost}). Indeed, for a given grouping with assignment matrix $X$, the assignment cost of $X$ is equal to (\ref{term_to_minimise}). The only difference with the classical problem is that $X$ requires to be symmetric.

Several efficient algorithms solve the general assignment problem. For instance, the Munkres' assignment algorithm or Hungarian method solves it in polynomial time \cite{3,4}. For a $n \times n$ cost matrix, the algorithm can achieve a $\mathcal{O} (n^3)$ time complexity. The implementation requires easy operations on the cost matrix. We use this method with the cost matrix defined previously to obtain a quasi-optimal grouping.

\section{Implementation issue}
Henceforth, we have an heuristic to compute the optimal matching. Given the cost matrix defined in (\ref{cost}), the Hungarian method will always find a solution. However the assignment matrix provided by the algorithm may not be symmetric, while our problem requires a symmetric $X$. We will now discuss the following two questions: (a) knowing that the cost matrix $C$ is symmetric, is there always a symmetric solution? (b) Can the Hungarian method find an optimal symmetric $X$ solution to our problem?  

To answer (a), we consider the following cost matrix
\begin{equation}
C = 
\begin{pmatrix}
3             & \underline{4} & \mathbf{1}\\
\mathbf{4}    & 7             & \underline{3}\\
\underline{1} & \mathbf{3}    & 2
\end{pmatrix},
\label{counterexample}
\end{equation}
where the bold and underlined numbers correspond to the two optimal assignments. We conclude that even if $C$ is symmetric, there is not always a symmetric assignment matrix that solves the problem. Thus \emph{the Hungarian method only provides an upper bound for the spectral efficiency}, where the bound equals the inverse of the assignment cost returned by the algorithm. 

Concerning (b), we will present how to use the Munkres' algorithm to obtain a quasi-optimal matching. Our simulations with the cost matrix defined previously show that the assignment matrix returned by the Hungarian method is generally not symmetric. This may be due to the fact that many coefficients in the matrix are the same. However, we remarked that for random matrices, the probability to obtain a symmetric assignment matrix is larger. Thus our idea is to add a small perturbation to the cost matrix $C$ and run the Hungarian method on the perturbed matrix $C'$. In practice, we compute $C'$ as follows: $C' = C + \epsilon$, where $\epsilon$ is a symmetric matrix whose coefficients are drawn according to $\mathcal{N}(0,\sigma^2)$ with $\sigma=10^{-3}$. 

Using this technique, we obtained a symmetric assignment most of the time (more details in the next section). We then compared the average spectrum efficiency of this grouping with the upper bound computed by the Hungarian method. As the results will point out, the performance of the matching obtained with the perturbed cost matrix is very close to the upper bound, justifying the term quasi-optimal of our method.

\section{Performance evaluation}
To evaluate the performance of the quasi-optimal matching, we use the simulation framework proposed in \cite{1,2}. We quickly give the main characteristics. More details can be found in the previous papers. First, the physical layer is based on the DVB-S2 standard \cite{5}. The code rates and modulations are resumed in Table~1. The hierarchical 16-APSK, which is not in the DVB-S2 standard, was introduced in \cite{1}. Then, the transmission is subject to additive white Gaussian noise (AWGN). Finally, the channel model to estimate the SNR distribution of the receivers in a spot beam takes into account two sources of attenuation: the relative location of the terminal with respect to the center of (beam) coverage and the weather. For a given simulation, the only parameter to set is the SNR at the center of the spot beam $SNR_{max}$.
\begin{table}[h]
\processtable{Transmission parameters}
{\renewcommand{\arraystretch}{1.05}
\begin{tabular}{l|c}\hline
Code rate  & 1/4, 1/3, 2/5, 1/2, 3/5, 2/3, 3/4, 4/5, 5/6, 8/9, 9/10 \\\hline
Modulation & QPSK, 8-PSK, 16-APSK, 32-APSK\\ 
           & hierarchical 8-PSK, hierarchical 16-APSK \\\hline
\end{tabular}}{}
\end{table}

\begin{figure}[h]
\centering
\includegraphics[width=0.975\columnwidth]{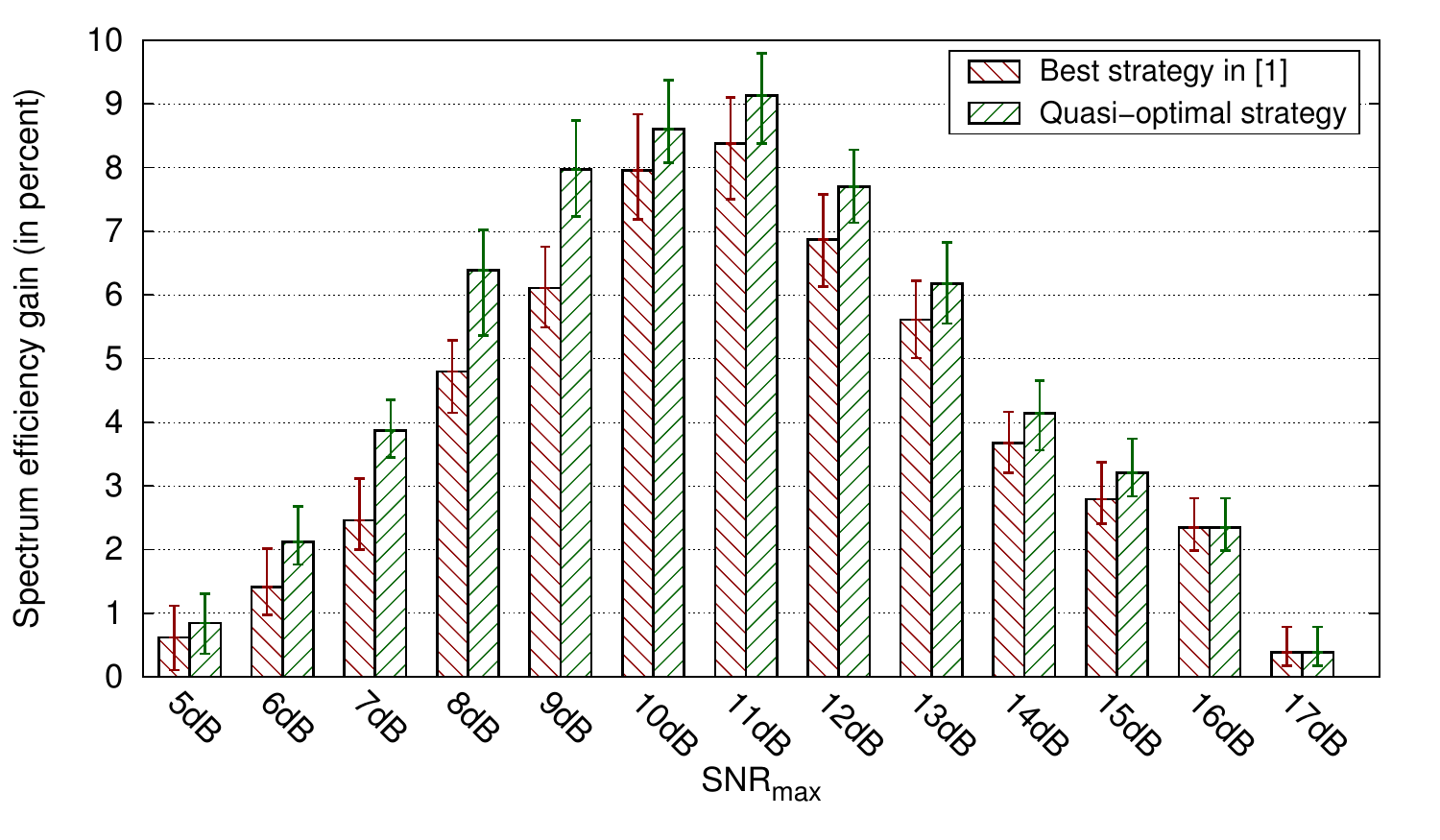}
\caption{Average spectrum efficiency gains (in comparison to time sharing without hierarchical modulation) with 500 receivers}
\label{results}
\end{figure}

\figurename~\ref{results} presents the gains (in terms of spectrum efficiency) when combining hierarchical modulation with time sharing for a broadcasting area with 500 receivers. For one system configuration (i.e., the parameter $SNR_{max}$ is set), we present the average, minimum and maximum gains over 100 simulations for the quasi-optimal matching obtained with the Hungarian method and for the best strategy in \cite{1} (which was described previously). The results point out that the quasi-optimal strategy provides some gains compared to the strategy proposed in \cite{1}. However the margin is small: in the best case ($SNR_{max}=9$ dB), we increase the performance from 6\% to 8\%. Even if we only notice a slight improvement, the maximum gain is now known for our framework.    

We mentioned before that the Hungarian algorithm does not generally return a symmetric assignment. To tackle this problem, we proposed to add a small perturbation to the cost matrix. During our simulations, our technique did not succeed to find a symmetric assignment for only a few cases: 8 simulations over 108 for $SNR_{max}=9$ dB and 4 over 104 for $SNR_{max}=10$ dB. Our method is thus quite reliable to find symmetric assignment. Moreover, we compared the performance of the symmetric assignment and the optimal assignment (provided by the Hungarian method and not necessarily symmetric). In all the simulations, the difference between the two is less than 1\%. In conclusion, we may not find the optimal symmetric assignment (as we use a perturbed cost matrix) but the performance is very close to the upper bound.  

Finally, we are interested to study the structure of the quasi-optimal groupings. Are the matchings random or is there a pattern (e.g., group the receivers with the largest SNR difference as in \cite{1})? To that end, we use the simulations to compute the probability that a coefficient in the assignment matrix $X$ is equal to one, i.e., $\Pr(X_{i,j}=1)$.

In order to see if there exists a pattern in the grouping process, we sort the receivers by increasing SNR. With that representation, the best strategy proposed in \cite{1} corresponds to the anti-diagonal matrix. Indeed, this strategy consists in grouping the two receivers with the largest SNR difference and repeating this operation. After sorting the receiver by increasing SNR, receiver $1$ which has the lowest SNR is matched to receiver $n$ which has the largest SNR, so $X_{1,n}=X_{n,1}=1$. For $1 \leqslant k \leqslant n$, receiver $k$ is matched to receiver $n+1-k$ and the corresponding assignment matrix is the anti-diagonal matrix. Note that time sharing, where there is no grouping, always corresponds to the identity matrix. 

\begin{figure}[!ht]
\centering
\subfigure[$SNR_{max} = 7$ dB]{
\includegraphics[width=0.475\columnwidth]{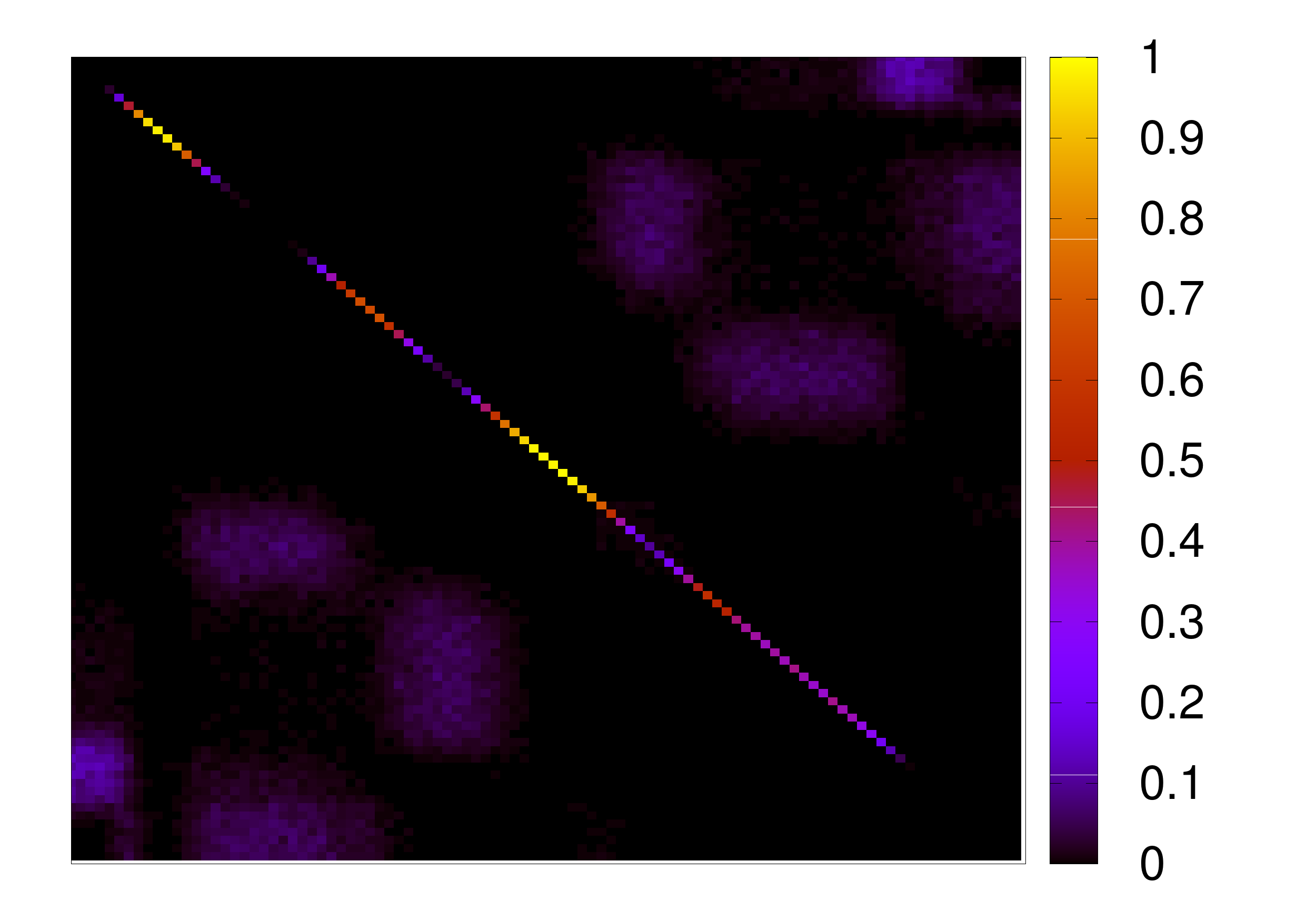}
\label{stats1}
}
\subfigure[$SNR_{max} = 10$ dB]{
\includegraphics[width=0.475\columnwidth]{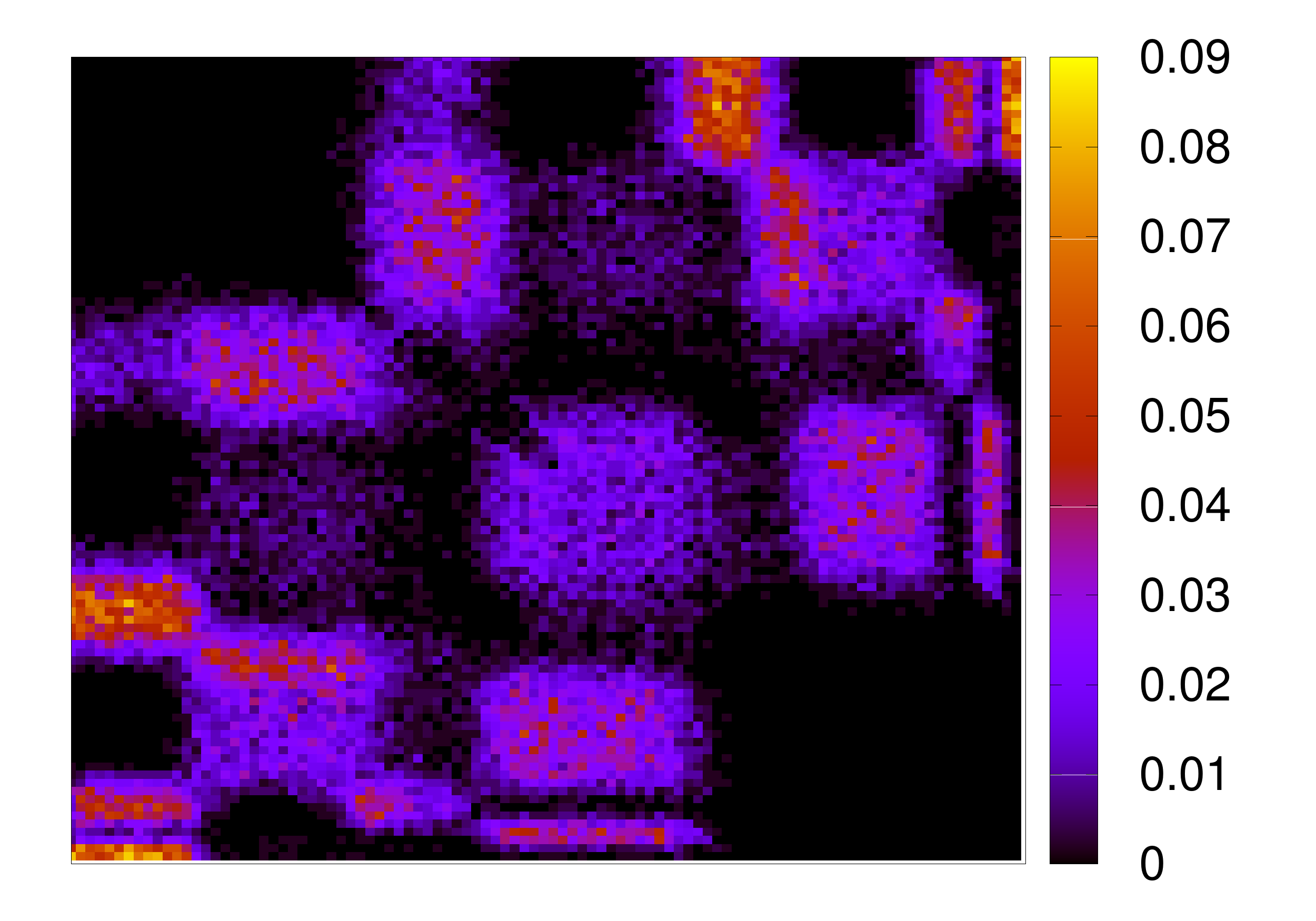}
\label{stats2}
}
\subfigure[$SNR_{max} = 13$ dB]{
\includegraphics[width=0.475\columnwidth]{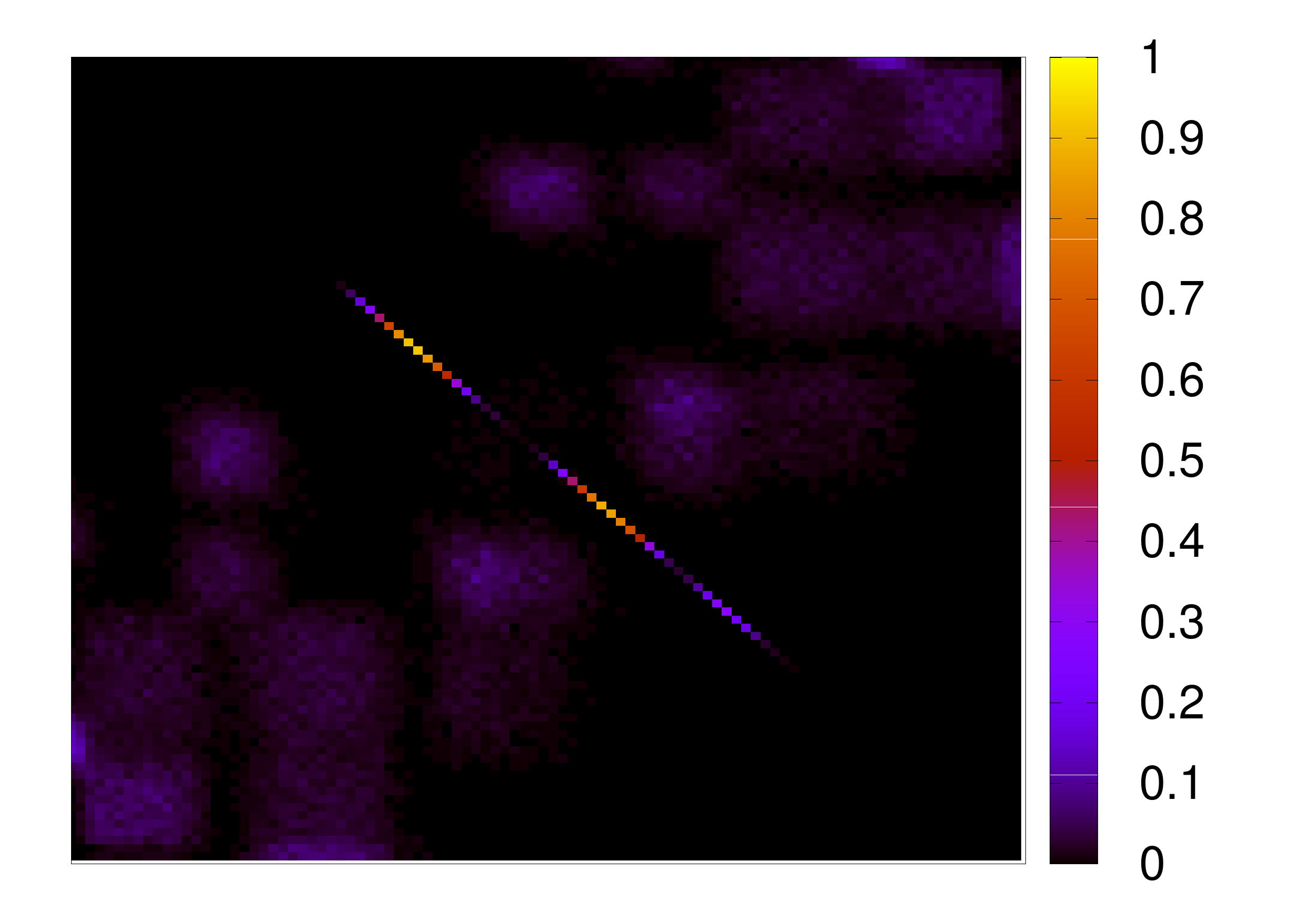}
\label{stats3}
}
\subfigure[$SNR_{max} = 16$ dB]{
\includegraphics[width=0.475\columnwidth]{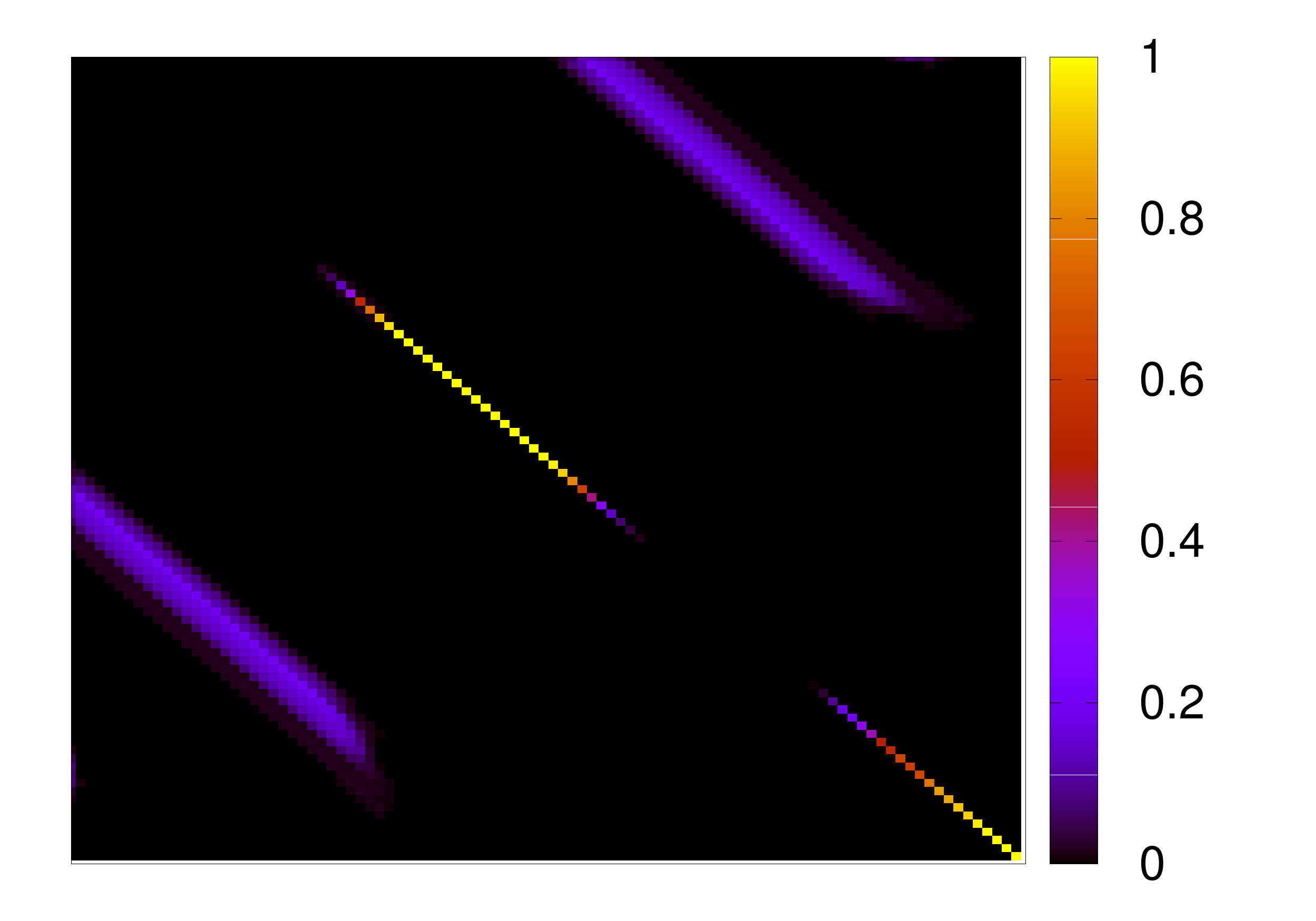}
\label{stats4}
}
\caption{Structure of the quasi-optimal assignment matrix ($\Pr(X_{i,j}=1)$)}
\label{pattern}
\end{figure}

We present the structure of the assignment matrix in \figurename~\ref{pattern} for several values of $SNR_{max}$. We see clearly in \figurename~\ref{stats2} and \figurename~\ref{stats3} that the quasi-optimal grouping usually matches low SNR receivers with large SNR receivers. This explains why the matching proposed in \cite{1} performs well. Moreover, when the gains compared to time sharing are low (see \figurename~\ref{stats1} and \figurename~\ref{stats4}), the optimal matching exhibits many single receivers as time sharing.

\section{Conclusion}
We showed how to obtain a close-to-optimal (in terms of spectrum efficiency) grouping strategy in a broadcast system that relies on time sharing and hierarchical modulation with two layers. The matching is the solution of an assignment problem that we solve using the Hungarian method. We compared the performance of the quasi-optimal strategy with a previously proposed matching and also studied its behaviour. In a future work, we will investigate how to obtain the optimal symmetric matching.
\vskip5pt

\noindent Hugo M{\'e}ric (\textit{INRIA Chile, Santiago, Chile})\\
\noindent E-mail: hugo.meric@inria.cl
\vskip3pt
\noindent Jos{\'e} Miguel Piquer (\textit{NIC Chile Research Lab, Santiago, Chile})
\vskip3pt
\noindent J{\'e}r{\^o}me Lacan (\textit{Universit{\'e} de Toulouse, Toulouse, France})
\vskip3pt



\begin{thebibliography}{}
\bibitem{1}
M\'eric, H., Lacan, J., Arnal, F., Lesthievent, G. and Boucheret, M.-L.: `Combining adaptive coding and modulation with hierarchical modulation in satcom systems', \textit{IEEE Transactions on Broadcasting}, 2013, \textbf{59(4)}, pp. 627-637
\bibitem{2}
M\'eric, H. and Piquer, J.M.: `DVB-S2 spectrum efficiency improvement with hierarchical modulation', \textit{IEEE International Conference on Communications}, 2014
\bibitem{3}
Munkres, J.: `Algorithms for the assignment and transportation problems', \textit{Journal of the Society for Industrial and Applied Mathematics}, 1957, \textbf{5(1)}, pp. 32-38
\bibitem{4}
Kuhn, H. W.: `The Hungarian method for the assignment problem', \textit{Naval Research Logistics Quarterly}, 1955, \textbf{2(1-2)}, pp. 83-97
\bibitem{5}
Morello, A. and Reimers, U.: `DVB-S2, the second generation standard for satellite broadcasting and unicasting', \textit{International Journal of Satellite Communications and Networking}, 2004, \textbf{22(3)}, pp. 249-268
\end{thebibliography}
\end{document}